\documentclass[pdflatex,sn-mathphys,a4paper]{sn-jnl}

\usepackage{caption}
\usepackage{subcaption}

\jyear{2024}%

\theoremstyle{thmstyleone}%
\theoremstyle{thmstyletwo}%

\theoremstyle{thmstylethree}%

\usepackage{xcolor}

\begin{document}

\title[Computationally accelerated materials characterization]{Computationally accelerated experimental materials characterization - drawing inspiration from high-throughput simulation workflows}

\author*[1]{\fnm{Markus} \sur{Stricker}}\email{markus.stricker@rub.de}
\author[2]{\fnm{Lars} \sur{Banko}}\email{lars.banko@rub.de}

\author[2]{\fnm{Nik} \sur{Sarazin}}\email{nik.sarazin@rub.de}

\author[3]{\fnm{Niklas} \sur{Siemer}}\email{siemer@mpie.de}

\author[3]{\fnm{Jan} \sur{Janssen}}\email{janssen@mpie.de}
\author[1]{\fnm{Lei} \sur{Zhang}}\email{lei.zhang-w2i@ruhr-uni-bochum.de}

\author[3]{\fnm{J\"org} \sur{Neugebauer}}\email{neugebauer@mpie.de}

\author[2]{\fnm{Alfred} \sur{Ludwig}}\email{alfred.ludwig@rub.de}

\affil[1]{\orgdiv{Interdisciplinary Centre for Advanced Materials Simulation}, \orgname{Ruhr-University Bochum}, \orgaddress{\street{Universit\"atsstr. 150}, \city{Bochum}, \postcode{44801}, \state{Northrhine-Westfalia}, \country{Germany}}}

\affil[2]{\orgdiv{Materials Discovery and Interfaces - Institute for Materials}, \orgname{Ruhr-University Bochum}, \orgaddress{\street{Universit\"atsstr. 150}, \city{Bochum}, \postcode{44801}, \state{Northrhine-Westfalia}, \country{Germany}}}

\affil[3]{\orgdiv{Department of Computational Materials Design}, \orgname{Max-Planck-Institut for Sustainable Materials}, \orgaddress{\street{Max-Planck-Stra\ss e 1}, \city{D\"usseldorf}, \postcode{40237}, \state{Northrhine-Westfalia}, \country{Germany}}}

\abstract{
Computational materials science increasingly benefits from data management, automation, and algorithm-based decision-making for the simulation of material properties and behavior.
Experimental materials science also changes rapidly by incorporation of `machine learning' in materials discovery campaigns.
The obvious benefits which include automation, reproducibility, data provenance, and reusability of managed data, however, is not widely available in the experimental domain.
We present an implementation of a Active Learning loop with a direct interface to an experimental measurement device in pyiron, a framework designed for high-throughput simulations, as demonstrator how to combine experimental and simulated data in one framework.
Apart from the acceleration provided by the active learning approach, additional acceleration of the experimental characterization is achieved by using prior knowledge from density functional theory simulations as well as composition-property predictions from literature mining using correlations in word embeddings.
With data from all domains in the same framework, a heretofore untapped and much-needed potential for the acceleration of materials characterization and materials discovery campaigns becomes available.
}




\maketitle

\section{Introduction}\label{sec1}

Computational materials science has developed from single calculations to high-throughput (HT) simulation campaigns in recent years.
This development is based on access to increased computational power and the development of simulation tools that allow for increasingly more complex simulation protocols.
But these capabilities present new challenges to manage calculations and the produced data.
In computational materials science, these challenges are addressed with integrated development environments such as pyiron~\cite{Janssen2019} or AiiDA~\cite{Huber2020} and other automation tools~\cite{Ong2013,Jain2015,Mathew2016,Mathew2017,Larsen2017,Oses2022}.
Major benefits of such approaches include automation, reproducibility, data provenance, and ensuring accessibility and reusability of existing data.

Experimental materials science is similarly shifting from manual, human-guided sample-by-sample synthesis and testing of individual samples towards HT (combinatorial) synthesis combined with HT characterization. Materials discovery in particular necessitates the screening of a practically unlimited amount of compositions and processing routes for specific functional properties~\cite{Ludwig2019}.
Samples in HT synthesis and characterization typically come in the form of composition spread materials libraries (CSML)~\cite{koinuma2004combinatorial}.
Conventionally, CSMLs are characterized by measuring chemical composition, structure, mechanical, optical, electrical, etc. properties by (semi-)automated measurement systems on a given number of pre-defined measurement areas, typically several 100s.
This degree of automation reduces the input of scientists and the time for measurements in characterization and analysis workflows.
Similarly, the analysis of the resulting relatively large datasets is challenging and often presents a substantial part of the total time needed for characterization. 
With several measurements on one CSML sample, combinatorial synthesis and HT characterization have similar challenges for data management~\cite{banko2020fast,Kraus2024} as HT simulation campaigns and the benefits provided by integrated development environments (IDEs) are equally desirable for experimental data and workflows.
Experimental data is often generated on distributed measurement instruments without or only limited metadata or data formatting standards.
And, similar to the approach taken for simulations, increasingly more automation and orchestration is incorporated into experimental materials science, e.g. BluesSky~\cite{Allan2019}, ChemOS~\cite{Roch2020}, or HELAO~\cite{Rahmanian2022}, including the benefits valid for  HT simulation campaigns when using integrated development environments.
A recent review of approaches can be found in~\cite{Benayad2022}. 
What is, however, still missing from all existing experiment-focused approaches is a direct interface to the computational domain.

Increasingly larger degrees of automation in experimental synthesis and characterization workflows provide the path to use the same optimization strategies employed in simulation workflows to accelerate the discovery and design of materials.
Once data from both domains exist in the same framework, leveraging the strengths of both approaches simultaneously becomes possible and provides further opportunities for acceleration.
Instead of triggering a calculation based on a suggestion provided by an optimizer, one can easily envision alternatively triggering a synthesis~\cite{coley2019robotic} and/or characterization routine to obtain the next (\textit{real}) material's properties in a discovery cycle~\cite{Kraus2024}.
In addition, latent knowledge from literature could also be seamlessly included as prior knowledge to predict material properties~\cite{Tshitoyan2019,Zhang2024}.
Ultimately, a global optimizer in a discovery cycle can then autonomously choose the \textit{best} next step w.r.t. cost, time, or uncertainty: either trigger a simulation or synthesis plus characterization of a real sample.
Further potential for optimization and acceleration lies in the possibility to efficiently access all previously measured and calculated properties and exploit existing knowledge to guide the optimization in the \textit{current} search space.

On the path to fuse data from simulations and experiments in one framework, we present a demonstrator using pyiron~\cite{Janssen2019} for experimental data acquisition.
Pyiron is an integrated development environment (IDE) for creating workflows~\cite{Janssen2019}.
It combines a job management system for automation and a hierarchical data management solution originally designed for atomistic modeling and is implemented in Python.
Its modularity allows to add custom jobs that make use of the built-in data management capabilities with minimal overhead.
Our demonstrator is designed to show that the concept of an IDE, developed for simulation workflows, can (a) be applied to experimental data acquisition, (b) directly optimize the measurement using an active learning strategy based on Gaussian process regression (GPR), (c) use existing data from simulations and word embedding-derived correlations of materials properties as priors, (d) analyze the data using automated data processing or data analysis routines~\cite{Maffettone2021, banko2021deep}, and (e) automatically solve the issue of data management and storage.
Our approach represents an extension and optimization of a standalone active learning approach for autonomous electrical resistance measurements~\cite{Thelen2023}.

Optimizing measurements in our context means decreasing the number of actual measurements while predicting/interpolating unmeasured regions in composition space with a defined uncertainty at the scale of the measurement uncertainty itself.
This directly translates into an acceleration of a characterization cycle, ultimately accelerating the whole materials discovery cycle.
Pyiron takes the role of an \textit{orchestrator} initializing the GPR with existing simulated data as a prior, running the active learning loop, querying the measurement device for the next optimal measurement point as well as storing and managing data and metadata of the process.
The application of GPR works because most composition-structure-property relationships show smooth trends that can be approximated by relatively simple functions of the composition, which is often the case for material systems that form solid solutions over a wide composition range, such as the class of materials commonly-called high entropy alloys~\cite{George2019,Xiao2022}.
Employment of an adaptive sampling strategy to approximate (instead of fully measure) a property as a function of the composition of CSMLs with a surrogate model based on a very small number of measurements results in an order of magnitude fewer measurements than brute-force automated approaches need as we show here.
The potential of integrated experimental-computational materials discovery has been demonstrated for organic materials~\cite{greenaway2021integrating}, nuclear materials~\cite{aguiar2020bringing,pyzer2022accelerating}, 
and is being considered to accelerate the discovery and optimization of urgently needed materials, e.g. for catalysis \cite{batchelor2021}.

Recent developments in the experimental field aim to integrate such workflows in automated, closed-loop discovery cycles, e.g. by using robotic platforms with integrated algorithms for data analysis and hypothesis generation~\cite{King2009,Kraus2024}.
It is envisaged that such platforms will enable fully autonomous research systems for materials discovery~\cite{burger2020mobile,kusne2020fly,Seifrid2022}.
Fully automated robotics platforms may be used in certain cases, however, it will be a long way to integrate several, especially high-quality in-depth characterization techniques (e.g. transmission electron microscopy, atom-probe tomography), and retrofitting existing equipment.
Nevertheless, it is possible and useful to create an integration of synthesis-characterization-evaluation cycles at the data level to unlock synergy effects between multiple data sources and to create partially (offline) autonomous research systems.
Ideally, computational and human resources should work collaboratively in such environments~\cite{Maffettone2021}.
The main difference between these systems and our approach, however, is the direct integration of simulations as well as latent knowledge from literature in the same framework as experimental measurements to benefit from all accessible knowledge to accelerate the materials discovery cycle.

Our demonstrator first and foremost shows that it is possible to include an experimental interface in pyiron and run a pre-defined workflow.
We further demonstrate that already existing data inside pyiron from density functional theory (DFT) simulations and text mining used as priors can be used to accelerate the active learning characterization loop.
This represents a step towards on-line data fusion from experiments, simulations and latent knowledge from literature with the potential to markedly accelerate materials discovery cycles.
The bigger picture is that IDEs such as pyiron have the potential to become integrated platforms for materials science with access to data from simulations and experiments and algorithm-assisted data analyses.

\section{Results}\label{sec2}

The demonstrator implementation of an active learning loop in pyiron communicates with an offline experimental dummy device that provides resistance measurements based upon query~\cite{Stricker2022c}.
The implementation has three ingredients: 1) pyiron as a basis to manage data, 2) a bespoke \texttt{ResistanceGP} ``job''\footnote{``job'' here is the pyiron-internal name for a defined workflow.} which controls the acquisition strategy, and 3) a custom interface to a measurement device.
pyiron itself is detailed in other publications~\cite{Janssen2019} and we, therefore, focus on the two other ingredients. A ``job'' in pyiron is a single calculation.
Concretely, the ``job'' here is an optimized measurement of the resistance on a CSML.
Its input comprises the user name, an identifier of the measurement device, a sample id, an initial parameter set including the maximum number of iterations for regression, and in our case a file path that points to a completed measurement (composition and resistance) to initialize the dummy measurement device which provides the actual \textit{measured} values upon request of the running job.
An overview is presented in Tab.~\ref{tab:pyiron_input_params}.
We want to stress that the choice is tailored to demonstrate the functionality and is not generic.
Since logging of inputs and outputs is done automatically, we can add any number of necessary or useful parameters automatically to a job through communication with a measurement device or a script triggering the characterization.
Ideally, each component communicates its settings at runtime, and all settings needed for the reproduction of the results are automatically saved along with the acquired data with minimal manual user intervention.
In the future, this could go as far as adding bar codes or radio-frequency identification (RFID) tags to samples, devices, and users for minimal manual intervention.

\begin{table}[h]
\caption{List of input parameters for the pyiron experimental job for optimized resistance characterization of the demonstrator.}
\begin{center}
\begin{tabular}{ll}
\toprule
Parameter & Description  \\
\midrule
\verb+exp_user+    & Name of the scientist\\
\verb+measurement_device+    & A unique name of the measurement device used\\
\verb+sample_id+    & Identifier of the physical sample\\
\verb+features+ & Input to the Regression\\
\verb+target+ & Output of the regression\\
\verb+initialization_indices+    & User-defined coordinates on CSML to initialize the GP\\
\verb+df+ & Dataframe for the dummy device\\
\verb+max_gp_iterations+ & Maximum iterations for GP\\
\verb+debug+ & Flag for saving intermediate results\\
\botrule
\end{tabular}
\end{center}
\label{tab:pyiron_input_params}
\end{table}

Within the ``job'', a custom interface to the measurement device is initialized and Gaussian process regression based on \texttt{GPy}~\cite{GPysince2012} is invoked.
Uncertainty sampling is then applied in which the algorithm determines the next measurement as a function of chemical composition which is translated into spatial x/y coordinates of the physical materials library.
The algorithm chooses the composition for which the model prediction has the highest uncertainty.
Five resistance measurements on the CSML are used to initialize the GPR.
This initial choice can be defined through user input but any known correlations from the composition-property space can serve as a prior without explicit user input.
A first example shows the use of an already existing dataset based on DFT calculations which correlates with the gradient of the resistance.
A second example demonstrates the initialization using a dataset of composition-resistance correlations created with a word embedding-based model~\cite{Zhang2024}.
The job subsequently requests the next measurement on the CSML by evaluating the composition for which the GPR predicts the highest uncertainty in predicting the target property resistance.
It runs until a stopping criterion is reached and the acquired data is automatically stored using the data management solution provided by pyiron.
Stopping criteria are not trivial as discussed in~\cite{Thelen2023}.
In our demonstrator we stop after a specified maximum number of iterations.

\begin{figure}[htp!]
    \centering
    \includegraphics{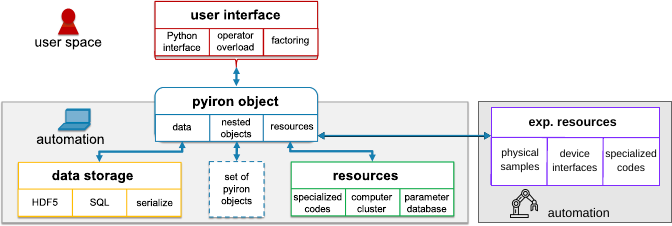}
    \caption{Internal structure of a pyiron object with the experimental extension on the right. The described extension to experimental workflows makes use of the existing infrastructure and is seamlessly integrated. Adapted from~\cite{Janssen2019} published under CC BY 4.0 license (http://creativecommons.org/licenses/BY/4.0).}
    \label{fig:pyiron_structure_exp_ext}
\end{figure}

From the perspective of pyiron, the only difference between a simulation ``job'' and a measurement ``job'' is the source of new data. The former source of data results from computational procedures and high-performance computing resources, the latter source of data is the output of a measurement device.
Fig.~\ref{fig:pyiron_structure_exp_ext} schematically shows the described extension to pyiron and its seamless integration to the environment.

\begin{figure}[htp!]
    \centering
    \includegraphics[width=\textwidth]{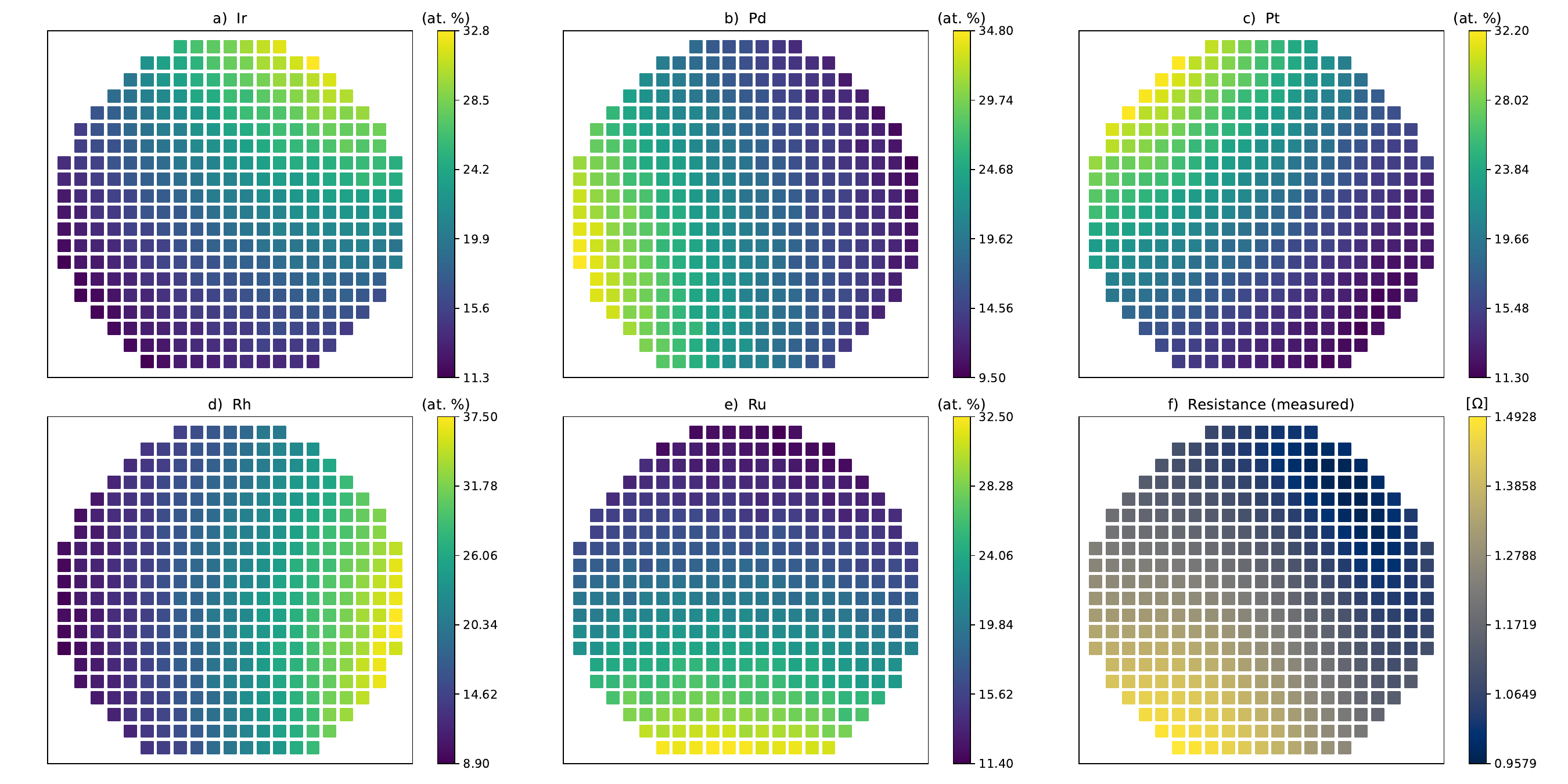}
    \caption{Chemical composition of the quinary noble metal system Ir-Pd-Pt-Rh-Ru CSML (a-e) in atomic \% and electrical resistance (f) for the 342 measurement areas.}
    \label{fig:quinary_noble}
\end{figure}

The composition and the measured electrical resistance of each point of the CSML used here for the demonstrator are shown in Fig.~\ref{fig:quinary_noble} and published via Zenodo~\cite{Banko2022a}.

Composition gradients are created by co-depositing from five pure elemental targets (Ir, Pd, Pt, Rh, and Ru) (cf. Fig.~\ref{fig:quinary_noble} a-e). Electrical resistance was measured by a four-point probe on a 342 point grid and is shown in Fig.~\ref{fig:quinary_noble} f.
The observed resistance trend is a smooth function of the composition, which motivates the use of an active learning scheme to reduce the number of measurements. 

\begin{figure}[htp!]
	\centering
	\includegraphics[width=.8\textwidth]{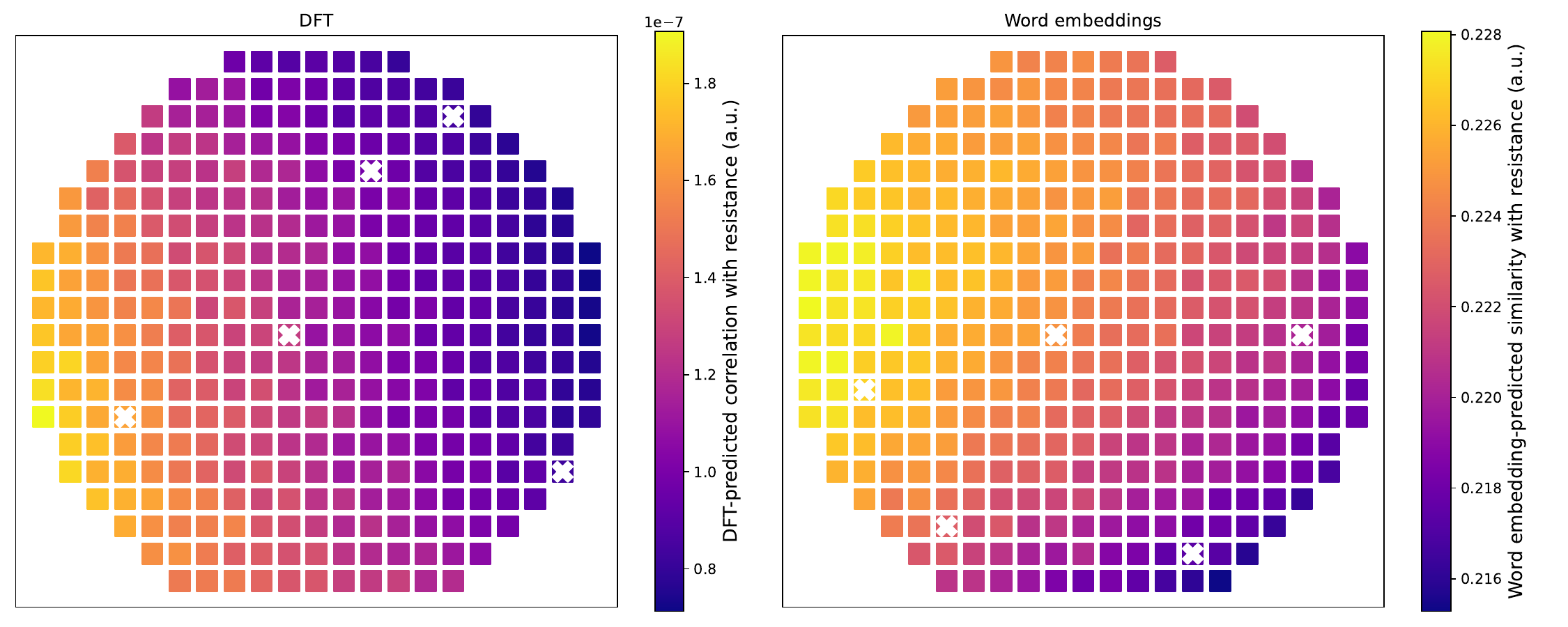}
	\caption{Composition-resistance correlations based on DFT calculations (left) and a word embedding-based model (left). Both are shown in arbitrary units as both models do not predict the resistance directly. White crosses mark the fastest converging initialization choices for a Gaussian process to approximate the data.}
	\label{fig:priors}
\end{figure}

The prior known correlations to be used for initialization of the active learning scheme are shown in Figure~\ref{fig:priors}.
On the left is a DFT-derived correlation based on the inverse of the Kubo-Greenwood formulation for the electrical conductivity for projector augmented wave sets~\cite{Calderin2017}.
Details of the implementation are provided in section~\ref{sec:methods} and the implementation can be found in~\cite{Janssen2024}.
On the right is a correlation with the resistance based on word embeddings using~\cite{Zhang2024}.
Details of the word embedding-based model are also provided in section~\ref{sec:methods}.
We intentionally chose a different colormap to indicate that the priors represent correlations with the resistance and not the actual resistance.
Both priors roughly approximate the gradient of the measured resistance (cf.Fig.~\ref{fig:quinary_noble}).
The maxima are approximately at the 9-o-clock positions, the minima at the 3-o-clock-position in contrast to the approximate diagonal positions of minimum and maximum of the measured resistance.

In a first step, we use both these priors and run 40 trial active learning loops starting with 5 different random points as initializations.
Since these runs are performed exclusively on the complete, existing data and do not require any experimental input they are very fast.
From each of these 40 random trials we choose the set of initialization indices which converge the fastest measured by the mean absolute error.
Figure~\ref{fig:priors} also shows the best initialization choices with white crosses.
These two initializations are then used as initializations of the GPR on the experimental data.
The idea is simple: we used already existing or cheap-to-compute correlations, find the best set of initializations for the actual measurement and thereby profit from prior knowledge about the composition-property relationship.
In machine learning terms we address the \textit{cold-start} problem of active learning~\cite{Houlsby2014,Zhu2020a,Barata2022}.

We compare these two choices based on prior knowledge with 40 random trials on the experimental data, a cross-shaped choice (6,9,12,3-o-clock positions), a choice based on the approximation of the DFT gradient (cross-shaped, max, min, center), and initializing the active learning loop at locations where each individual element has its maximum in the composition (cf.~\ref{fig:quinary_noble} for the maxima).

\begin{figure}[htp!]
    \centering
    \includegraphics[width=\textwidth]{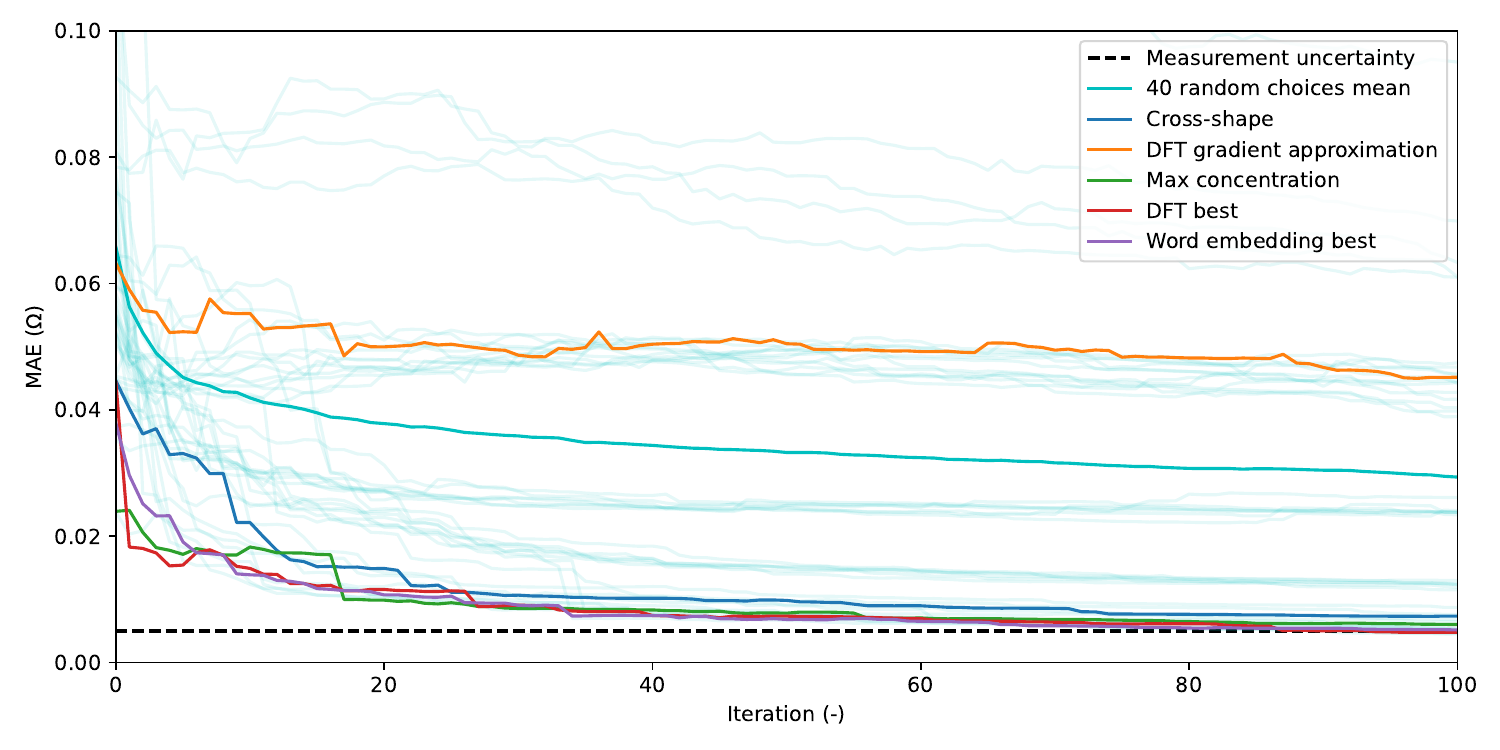}
    \caption{Mean absolute error (MAE) evolution for active learning loops with various initialization choices. The measurement uncertainty of the method ($\pm 0.005\,\Omega$ at approx. $1\,\Omega$) is indicated with a dashed black line. Faint cyan lines represent 40 different random initializations; their average is the dark cyan line. Initialization with a cross-shape (6,9,12,3-o-clock) in blue, approximation of the DFT correlation gradient in orange, initialization of the compositions with individual element concentration maxima in green, best initialization choice from randomly-initialized DFT and word embedding-based active learning loops in red, resp. purple.}
    \label{fig:mae_vs_iteration_all}
\end{figure}

Fig.~\ref{fig:mae_vs_iteration_all} shows the evolution of these initialization choices as well as the measurement error of the order of $\pm 0.005\,\Omega$ at approx. $1\,\Omega$ (cf. Figure~\ref{fig:quinary_noble}, bottom right).
Dark cyan represents the average of the 40 random initializations which are shown in faint cyan.
This serves as our reference: any faster convergence than the average random choice is considered ``accelerated''.
In blue, the convergence of the cross-shaped initialization is shown.
It converges relatively fast.
Orange shows an initialization based on the gradient of the DFT covering the maximum and minimum, the center as well as two points orthogonal to that line.
Convergence is not observed.
The green line shows the initialization at the maxima of elemental concentrations.
Convergence is fast.
The \textit{best} DFT- and word embedding-derived initialization choices (Fig.~\ref{fig:priors}) in red and purple show very fast convergence.
Particularly the word embedding-derived initialization is within the measurement error at slightly over 50 iterations.
This is where the an active learning loop could stop without sacrificing much of the accuracy w.r.t. the full measurement of the CSML (cf.~\cite{Thelen2023} for a discussion about stopping criteria).

\section{Discussion}\label{sec12}

The presented example of a pyiron-controlled experiment with a dummy experimental device in the loop shows the potential for future applications.
Instead of brute-force measuring the resistance of all 342 compositions provided by the discretization of the materials library, only approx. 50 compositions need to be measured and the rest is interpolated including the uncertainty for the interpolation.
Consequently, the saved characterization data of the CSML includes 50 actual measurements and 292 interpolations (predictions) that are marked as such including the model used for interpolation, translating into an acceleration of the process of almost one order of magnitude (7-fold).
A typical time to obtain one measurement is $\approx 10\,$s including moving the measurement tip to the position. I.e. the measurement of one CSML with 342 measurement points is about one hour. Measuring only $50$ points translates to $\approx 8.5\,$min.
We state this matter-of-factly but the consequences on how existing materials data could be used for accelerated characterization in the near future need to be stressed.
In the context of materials discovery for combinatorial problems, this is the required quality of acceleration which enables faster screening of materials properties.

But the real benefit lies in the central storage and reusability of the data as priors.
We have reused already existing data (DFT and word embedding correlations) to initialize, and therefore accelerate, the active learning loop.
Compared to the average random initialization (Fig.~\ref{fig:mae_vs_iteration_all}, dark cyan line), the active learning loops including prior knowledge at \textit{cold-start} reach a desired mean absolute error within approx. 50 iterations for this CSML.
A stopping criterion following~\cite{Thelen2023} could additionally be included.
In~\cite{Thelen2023}, the authors discuss the need for a stopping criterion which leads to generally more measurements given that the active learning loops were initialized with 9 starting points roughly covering the sample.
We expect that the usage of priors to adapt the initialization would accelerate convergence.
However, for our demonstrator, we had to create the priors in the first place.
The DFT calculations require many intensive calculations: 10 structures with 108 atoms per cell per each of the 342 compositions.
This is an unlikely scenario because both, the required time as well as the energy spent are hardly justifiable to accelerate the characterization of a real sample.
The word embedding-based prior, however, is cheap and predictions for new compositions could be triggered automatically.
In our description and for our argument we assume that both priors existed.
This scenario seems more likely in the future: we think that in the future we will see extensive hybrid computational/experimental data bases that are queryable on demand by automated workflows.

\begin{figure}
\begin{subfigure}[t]{0.45\textwidth}
    \centering
    \includegraphics[width=\textwidth]{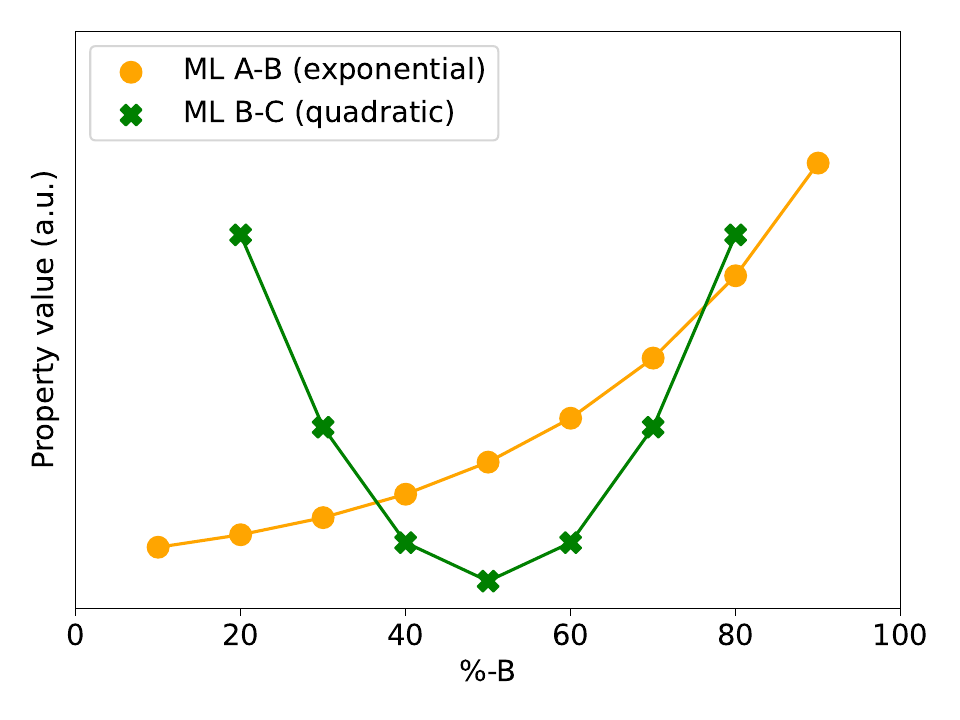}
    \caption{Schematic example property measurements of two materials libraries with elements A-B and B-C.}
    \label{fig:AB-BC-property}
\end{subfigure}
\hfill
\begin{subfigure}[t]{0.45\textwidth}
    \centering
    \includegraphics[width=\textwidth]{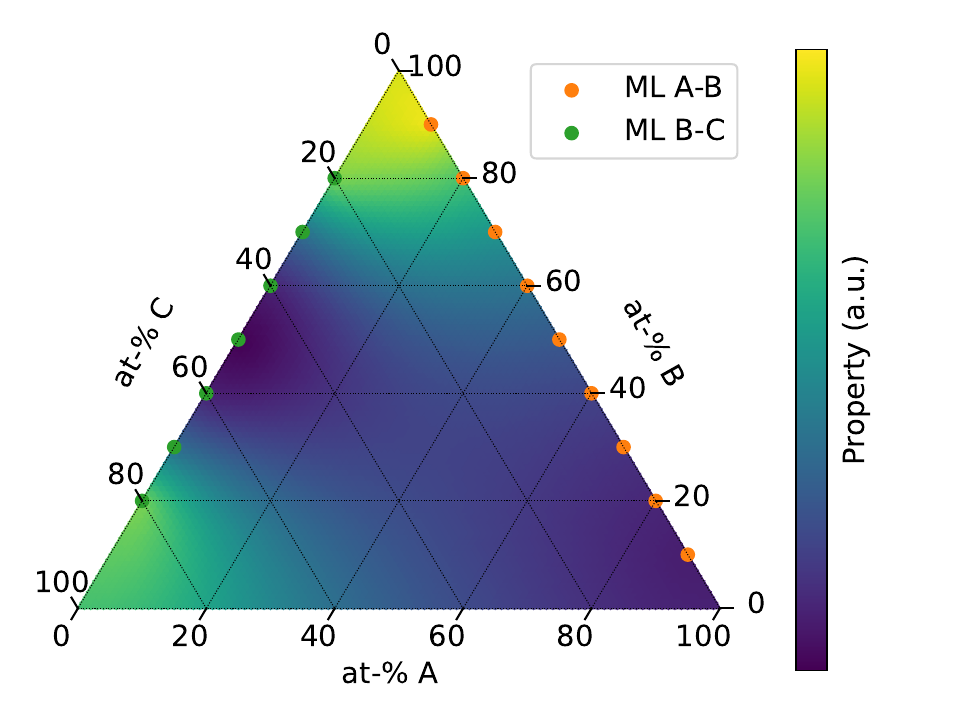}
    \caption{Schematic of a ternary search space with \textit{edge} data (A-B and B-C) already measured and the rest interpolated.}
    \label{fig:two_binary_to_ternary_illustration}
\end{subfigure}
\caption{Schematic property measurement (a) for two binary systems with each two elements A-B and B-C shown both as a function of the elemental content of element B. The property measurement on the library A-B shows an exponential dependence on B-content, B-C is quadratic for two binary systems. And (b) the possible property prediction on the ternary prior to any measurements. The characterization of the \textit{full} A-B-C compositional space can be initialized with the existing dataset: The green and orange points would substitute the initial choice of measurements (marked with white crosses) in Fig.~\ref{fig:priors}. This concept is directly applicable to material systems with more elements. A ternary was chosen solely for presentation purposes.}
\label{fig:two_binary_ternary_schematics}
\end{figure}

Based on this expectation, imagine a discovery campaign is started in a compositionally (partially) overlapping material system.
Any overlap with a previous measurement potentially accelerates the characterization of new systems (predicted or actually measured).
E.g. in~\cite{Zhang2024a}, the authors follow exactly this: based on existing measurements of two overlapping ternary CSMLs, they fit models to successfully predict composition-property relationships of the shared quaternary CSML.
In the future, unknown regions of a composition-property space could even be explored autonomously without the initial need for measurements to identify promising regions first.
Autonomously here refers to an automatable procedure in the combined research data-computational framework which routinely checks overlaps in compositions of stored measurements to automatically predict property behavior.
A schematic of such an overlap is shown in Fig.~\ref{fig:AB-BC-property}.
Two materials libraries exist in the database, one containing a property measurement of combinations of elements A and B, a second one containing B and C.
The overlap is given by element B.
These two measurements in two binary systems comprise the edges of a hypothetical ternary system A-B-C.

With an increasing number of measurements stored in the same accessible database, the larger this benefit becomes and actual sample preparation can then target ``very high uncertainty'' regions.
A possible pitfall of the models for predicting unmeasured properties is that they all might use different local coordinate systems of elements.
In our demonstrator the model is

\begin{equation}
    R = f(c_{\textrm{Ir}}, c_{\textrm{Pd}}, c_{\textrm{Pt}}, c_{\textrm{Rh}}, c_{\textrm{Ru}}),
\end{equation}

where $R$ is the resistance and $c_{i}$ the respective compositions.
But a GPR, as presented here, can easily be adapted to different element-based coordinate systems by treating the measured and interpolated known values as given, adapting the coordinate system, and setting any additional element $c_{\textrm{add}}$ to zero:
\begin{equation}
    R = f(c_{\textrm{Ir}}, c_{\textrm{Pd}}, c_{\textrm{Pt}}, c_{\textrm{Rh}}, c_{\textrm{Ru}}, c_{\textrm{add}} = 0)
\end{equation}

to fit a model.
With such a model, a complete edge or multiple edges of new compositional spaces can be used as priors.
Fig.~\ref{fig:two_binary_to_ternary_illustration} shows this schematically for the individual two measurements presented in Fig.~\ref{fig:AB-BC-property}.
Two CSML with elements A-B and B-C are present as data sets, the characterization of the ternary A-B-C can then be initialized with known properties on the edges.
This could even be done automatically. E.g. if samples are tagged with an RFID chip, any existing characterization data and possible priors like energy dispersive X-Ray (EDX) measurements could be looked up in the database. If EDX data is present, this can be used as a prior in GPR, if not, some other sample positions are suggested to be measured first to initialize the GPR.
An alternative to using the composition as a representation for ``a material'' are representations of materials based on word embeddings~\cite{Zhang2024a}.
These are fixed-size vectors which allow simpler screening for high-dimensional compositions spaces, however, they depend on the corpus on which they are based.

If characterization data from less complex binary and ternary materials from CSMLs are present, more property predictions for materials with more constituents can be assembled.
Given sufficient data and limited expectations about the certainty of a prediction/interpolation, further measurements could even be unnecessary to explore unknown composition spaces.
In all cases, this requires strict bookkeeping~\cite{Dudarev2024} where material properties originated (measurement, prediction, simulation) and propagation of uncertainty for which frameworks like pyiron are ideal.
The commonly-used term in research data management for this is \textit{provenance}.

Another possibility for informed initialization (``improving the prior'') without existing data is to directly trigger simulations or word embedding predictions from an experimental job, which provide a first approximation of the composition-property relationship to be measured and use that as a prior as presented above.
More related data available prior to any measurement provides a possibility to accelerate the active learning loop.
However, the possibility to either trigger simulations or other cheap proxy experiments prior to expensive measurements only exists if both domains are unified in one framework.

Before this can become a standard, several technical, as well as \textit{acceptance} challenges need to be overcome.
Practical technical challenges include that many experimental devices are controlled by proprietary software without APIs, effectively rendering them unusable in the outlined approach.
Solutions to this issue are obvious:
Either a vendor provides an API through which a device can be controlled or an open-source operating system is run to control the device directly.
We expect that this will be solved with time.
Operationally, however, and this is a drastic change in how data and data management will be seen: researchers need to be aware and accept that not every data point in the database is actually measured, but the majority of them will be interpolations/predictions.
In our solution, a surrogate model provides most of the data: $\approx 6/7$ values.
This is in part a communication challenge but in our view a larger part an issue of trust in software solutions, particularly in the context of machine learning ideally producing interpretable models~\cite{Oviedo2019}.
Our model as a mathematically correct concept and, even more important, its correct implementation needs to be ensured.
The former is a theoretical problem, which is in principle solved, the latter requires strict standards for implementation through unit tests, continuous integration tools, and benchmarking during software and method development~\cite{Dunn2020,Rohr2020b,Haese2021,Stein2022a}.

The last point shows a current gap between the simulation and experimental materials science domains.
Computational materials scientists are used to using (legacy) code and routinely test and trust software.
Surrogate models, their implementation in computer code, and computational resources are the \textit{tools of the trade}.
Many computational materials scientists go through formal or informal training for coding and software development along with their materials science training which allows them to check their own as well as others' codes and their physical validity.
Researchers trained in experimental sciences also undergo a formal training, e.g. in sample preparation and device control, but usually, only a few have the necessary background to assess, use, and change codes and use computational resources which in turn can result in less or no trust in predicted results.
But the solution for this is also clear: as the experimental and simulation domains are merged on the level of data, the formal (or informal) training of researchers from different sub-disciplines also needs to converge.
Experimentalists who use these tools and frameworks need to be trained.
And while the initial overhead of time spent for training might be seen as a drawback, the potential for acceleration of this approach in the context of materials discovery is real and becomes larger as more data is collected in a unified framework.
Further, recent developments in lab automation~\cite{Seifrid2022,MacLeod2022,pyzer2022accelerating} will lead to less training spent to operate devices.
Adding a new experimental data point will then ideally be as simple as running a simulation, the call of a function within a framework like pyiron.

In summary, the use of a unified framework as outlined puts less importance on how data is acquired (simulation, measurement, prediction) and more importance on how data and metadata are stored and that they can be reused in subsequent research.

\section{Conclusion}\label{sec13}
We demonstrate that an integrated development environment designed for use with high-throughput simulations is also able to control an experimental procedure, similar to triggering a calculation on a high-performance computing platform.
With this approach, the boundaries between data obtained from experiments or simulations vanish because \textit{data} does not care about its source.
This approach allows to leverage computational power through the usage of existing priors to accelerate materials characterization, here demonstrated by using Gaussian process regression to reduce the number of measurement points by approximately one order of magnitude while predicting non-measured points with the desired uncertainty.
By embedding experimental workflows in pyiron, we automatically benefit from automation, reproducibility, and data provenance, and at the same time ensure accessibility and reusability of existing data.
Accessibility and reusability are key ingredients for further acceleration of characterization because an existing database provides prior knowledge to cold-start, e.g., a Gaussian process more strategically instead of manually or randomly.
This is the first step towards automatically fusing data from different sources, which allows a composition-property-space-level optimization of materials discovery campaigns.
Necessary changes in software and hardware accessibility for this vision to become a reality are APIs to experimental devices and training of researchers to create, trust, and use workflow managers for data acquisition.

Ultimately, we envision a (software) layer in which a federation of algorithms (sometimes also referred to as ``agents'') use results obtained by other agents as prior information for autonomous characterization, thereby lowering the number of needed measurements to ultimately speed up characterization time without compromising (too much) on data precision or sampling.
The final characterization of a given material is then a mix of actual measurements and predictions each with its associated uncertainty.
Reproducibility and confidence in data is ensured because each step of the procedure is automatically documented by the workflow manager.

\section{Methods}\label{sec:methods}

\textbf{pyiron extension}
The extension of pyiron comprises the following parts.
First, we added a new `job` that inherits the basic functionality and integration to the installed pyiron instance automatically.
Necessary additions include the definition of user-defined input variables, a recipe for the Gaussian process-based measurement, and functionality to retrieve the measured as well as the interpolated data along with the model.
The input variables currently comprise a user, a measurement device name, a sample ID, initial measurement points for the Gaussian process, a sample file, and the maximum number of iterations (Tab.~\ref{tab:pyiron_input_params}).
In a future iteration, this user input will serve as a basis for the automatic generation of metadata for datasets and will grow and change over time as more devices and measurement routines are implemented.
The version used in this paper is published here~\cite{Stricker2022b} along with an example notebook on how to use it and to reproduce the figures of this article.

\textbf{Experimental dummy device and wrapper for GPy}
The experimental dummy device simulates an API to an experimental setup and is implemented as a simple class in Python and can be found in a utility repository~\cite{Stricker2022c}.
Its API mimics the call to an actual device and returns the desired property (here: resistance) of a certain composition on a materials library value.
This API to a dummy device will be replaced by an API to actual experimental setups once they become available.
Additional to the dummy device, a wrapper for the Gaussian processes framework GPy~\cite{GPysince2012} is also included here.
The wrapper is a class that allows convenient initialization, update, and prediction of a Gaussian process as implemented in `GPy'.

\textbf{Density functional theory calculation details}
The density functional theory (DFT) calculations are performed using the Vienna ab-initio simulation package (VASP)~\cite{Kresse1993,Kresse1994,Kresse1996a,Kresse1996b} in version 5.4.4, with the extension to calculate Kubo-Greenwood transport properties~\cite{DiPaola2020a}.
The plane wave energy cut-off is set to 500eV, in combination with a Gamma-centered 3x3x3 k-point mesh and Fermi smearing of 25meV.
To simulate the chemical complexity of the CSML in a periodic simulation cell, 108-atom FCC special quasi-random structures (SQS) are generated using the sqsgenerator software~\cite{Gehringer2023}.
The lattice constants are interpolated by the rule of mixture based the end-member nearest neighbor distance as retrieved from the materials project~\cite{Jain2013}.
For each of the 342 experimental compositions 10 SQS structures are generated to calculate the electrical resistivity by averaging. The resulting composition-resistance correlations are shown in Fig.~\ref{fig:priors}.
The pyiron~\cite{Janssen2019} workflow used to perform the DFT prediction of the resistivity of a given CSML is available on Github~\cite{Janssen2024}.
The workflow is based on the VASP ``job'' class, starting with a parallel calculation to optimize the electronic structure.
It is followed by a serial calculation to store the electronic structure for post-processing.
Finally, a new ``Kubo-Greenwood'' postprocessing ``job'' class is implemented to calculate the transport properties.
To enable the high-throughput screening of a wide-range of composition-resistance correlations the modelling of the resistance uses static calculations, i.e. without relaxing the box shape or the atomic positions or performing molecular dynamics calculations for averaging the electronic transport properties.
We note that the implementation in pyiron is flexible and can be straightforwardly extended to include additional approximations. 

\textbf{Word embedding model details}
The correlation between word embeddings and resistance for was evaluated using a trained word2vec model. We used the \texttt{VecGenerator} module in the MatNexus~\cite{Zhang2024}.
To create a high-quality word2vec model, abstracts related to electrocatalysts and compositionally complex solid solutions were collected through the \texttt{PaperCollector} module from databases such as Scopus and Arxiv. The query targeted relevant research papers available in open-access format up to the year 2024. 
In the preprocessing stage, abstracts were refined through multiple steps in the \texttt{TextProcessor} module.
Filtering comprises typically unwanted characters, including copyright symbols (e.g., “\textcopyright”) and references to legal entities (e.g., “\& Co.”).
The text was then tokenized into individual words, and common English stopwords (e.g., “and,” “the”) were filtered out.
For consistency, words were lemmatized to their root forms.
Additionally, chemical symbols were explicitly identified and retained in the final corpus.
For model training, the skip-gram algorithm was selected as it effectively captures rare word associations.
We used an embedding dimension of 200, a context window of 5 words, and hierarchical softmax was applied to optimize training efficiency.
A minimum word count of 1 was set, ensuring all terms in the specialized corpus contribute to the learned embeddings.
The trained word2vec model was then used to predict correlations with experimental resistance values from for the used dataset containing elements Ir, Pd, Pt, Rh, and Ru.
Using the \texttt{MaterialSimilarityCalculator} within \texttt{VecGenerator}, we used the model to predict correlations with resistance by assessing the  similarity scores between word embeddings for the compositions in the CSML and the word `resistance'.

\textbf{Snythesis and characterization of composition spread materials libraries} 
The CSML was fabricated by sputter co-deposition from five pure elemental targets.
Fabrication details are provided in~\cite{banko2022unravelling}. 
Electrical resistance was measured using an automated four-point probe test stand.
The setup is described in~\cite{thienhaus2011modular}.
Chemical composition was measured by energy-dispersive X-ray spectroscopy in a scanning electron microscopy Jeol 5800 LV equipped with an Oxford X-act detector. The acceleration voltage was set to 20 kV.
The final data set of CSML used in the dummy experimental device with chemical characterization and resistance measurements is published on Zenodo~\cite{Banko2022a}.

\textbf{Data availability}
The experimental data used in the demonstrator is published on Zenodo~\cite{Banko2022a}.
Summarizing tables for the DFT and word embedding-based as well as the measurements are also part of the project repository~\cite{Stricker2022b}.

\textbf{Code availability}
The code used in this study is available here~\cite{Stricker2022b}. The implementation of the experimental dummy device is available here~\cite{Stricker2022c}.
Pyiron workflows for the DFT studies are available here~\cite{Janssen2024}.
The word embedding model used here can be found on~\cite{Zhang2024b}.

\textbf{Author contributions}
MS, LB, and AL designed the study. NSa carried out the measurements and developed a first prototype of the Gaussian process regression. MS formalized the prototype and developed the utility library. MS, JN, LB, and NSi implemented the custom pyiron job. LZ and MS developed the word embedding model and predicted the similarities. JJ and JN performed the DFT calculations. All authors analyzed the results and wrote the manuscript.

\textbf{Competing interests}
The authors declare no competing financial or non-financial interests.

\backmatter

\bmhead{Acknowledgments}
AL gratefully acknowledges funding from Deutsche Forschungsgemeinschaft (DFG) through project LU1175/26-1.
LZ and MS gratefully acknowledge the financial support provided by the China Scholarship Council (CSC number: 202208360048).
MS, LB, JN, and AL gratefully acknowledge funding by Deutsche Forschungsgemeinschaft (DFG) -- CRC1625, project number 506711657, subprojects A01, A05, INF.

\bibliography{literature}

\end{document}